\documentclass[twocolumn,prb,showpacs]{revtex4}

\usepackage{graphicx}
\usepackage{epsfig}
\usepackage{dcolumn}
\usepackage{bm}
\usepackage{amsmath}

\begin{document}
\title{Spin blockade at semiconductor/ferromagnet junctions}
\author{Yuriy V. Pershin}
\email{pershin@physics.ucsd.edu}
\author{Massimiliano Di Ventra}
\email{diventra@physics.ucsd.edu} \affiliation{Department of
Physics, University of California, San Diego, La Jolla, California
92093-0319}

\begin{abstract}
We study theoretically extraction of spin-polarized electrons at
nonmagnetic semiconductor/ferromagnet junctions. The outflow of
majority spin electrons from the semiconductor into the ferromagnet
leaves a cloud of minority spin electrons in the semiconductor
region near the junction, forming a local spin-dipole configuration
at the semiconductor/ferromagnet interface. This minority spin cloud
can limit the majority spin current through the junction creating a
pronounced spin-blockade at a critical current. We calculate the
critical spin-blockade current in both planar and cylindrical
geometries and discuss possible experimental tests of our
predictions.

\end{abstract}


\maketitle

The use of electron spins in semiconductors and their dynamics
across semiconductor/ferromagnet (S/F) interfaces shows great
promise for device applications~\cite{book,review,r19,datta}. Most
of the theoretical and experimental attention so far has been
focusing primarily on mechanisms of spin injection from the
ferromagnet to the semiconductor, spin transport and spin relaxation
in semiconductors
\cite{Flatte,r1,r2,saikin,bleibaum,rashba,pershin,r3,r4,r5,r6,r7}.
However, it is believed that a functional spintronic device
\cite{datta} would not only involve injection of spin-polarized
electrons from the ferromagnet to the semiconductor, but also the
reverse process: the {\em extraction} of spin-polarized electrons
from the semiconductor to the ferromagnet. Despite the apparent
similarity with the injection process and recent experimental and
theoretical progress in this area
\cite{r8,r8a,stephens,crooker,sham,sham1}, the physics of spin
extraction has not been fully explored yet.

The main experimental breakthrough in this field is the discovery
\cite{r8,r8a} and observation \cite{stephens,crooker} of the
ferromagnetic proximity effect in several systems. In these
experiments, a spontaneous electron spin polarization of several
percents in magnitude has been generated optically and
electronically in the vicinity of the interface in the semiconductor
region. An interesting finding is that the direction of spontaneous
spin polarization can be parallel or antiparallel to the
magnetization of the ferromagnet. These experiments have been
explained using scattering theory \cite{sham} and its extension
\cite{sham1}. According to this theory, spontaneous spin
polarization near the interface appears because spin-up and
spin-down electrons have different probabilities to enter into (or
to be reflected from) the ferromagnet.

In this paper, we consider spin extraction from a nonmagnetic
semiconductor with a non-degenerate electron gas into a
ferromagnet in the regime when the degree of spontaneous spin
polarization near the interface is high (close to 100$\%$). We
show that the most important feature of this regime is that the
cloud of spontaneous spin polarization of minority spins limits
the majority spin current through the junction via a spin blockade
when a critical current is reached. We discuss this phenomenon at
both planar S/F interfaces and at an interface between a
semiconductor and a ferromagnet of cylindrical shape. The latter
case is relevant for scanning tunneling microscopy (STM)
experiments with ferromagnetic tips. We show that the
spin-blockade of the current is more important in materials with
long spin relaxation times. Therefore, this novel phenomenon is
fundamentally relevant for the design of future spintronic devices
and can be readily verified experimentally.

\begin{figure}[b]
\includegraphics[angle=270,width=7.5cm]{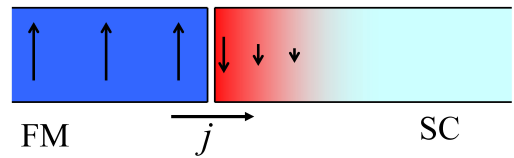}
\caption{\label{fig1} (Color online) Schematic of spin
polarization distribution in the biased semiconductor/ferromagnet
junction. The flow of spin-up electrons from the semiconductor
(SC) into the ferromagnet (FM) results in higher concentration of
spin-down electrons near the junction.}
\end{figure}

Let us start to discuss the planar S/F interface. Figure~\ref{fig1}
shows the system under investigation consisting of a junction
between a ferromagnetic material and an $n$-doped nonmagnetic
semiconductor. We assume that a bias is applied to the system in
such a way that the electron flow is directed from the nonmagnetic
semiconductor into the ferromagnet. The electrons incoming from the
bulk of the semiconductor are spin-unpolarized. Let us start by
considering a perfect ferromagnet, such as a ferromagnetic
half-metal. The latter accepts only (say) spin-up electrons at the
junction. Therefore a cloud of spin-down electrons (which can not
enter into the ferromagnet without undergoing spin reversal) must
form in the semiconductor side in proximity to the junction (see
Fig. \ref{fig1}). It is obvious that the cloud of spin-down
electrons increases with current.  The spin blockade occurs then at
a certain current magnitude, when the semiconductor region near the
junction becomes completely depleted of electrons having the same
direction of spins as the majority spin electrons in the
ferromagnet.

For our analysis of this phenomenon the detailed structure of the
interface is not very important. We therefore solve spin transport
equations for the semiconductor region only. The calculations are
performed at a fixed current through the structure. Using the
current as the external control parameter, rather than the applied
voltage, is more convenient because the current is constant
throughout the electric circuit that contains the sample. On the
other hand, if we use the voltage as the external control parameter,
we have to take into account voltage drops in different parts of the
circuit, such as, for example, at the Schottky barrier between metal
and semiconductor. The junction with the ferromagnet is then taken
into account through the boundary condition on current components by
neglecting space charge effects. The critical current is found from
the condition of zero spin-up density at the junction.

Our theory is based on the two-component drift-diffusion model
\cite{Flatte}. The system of drift-diffusion equations consists of
the continuity equations for spin-up and spin-down electrons and the
equations for the current:

\begin{equation}
e\frac{\partial n_{\uparrow (\downarrow)}}{\partial
t}=\textnormal{div} \vec j_{\uparrow
(\downarrow)}+\frac{e}{2\tau_{sf}}\left(n_{\downarrow
(\uparrow)}-n_{\uparrow (\downarrow)} \right), \label{contEq}
\end{equation}
\begin{equation}
\vec j_{\uparrow (\downarrow)}=\sigma_{\uparrow (\downarrow)}\vec
E+eD\nabla  n_{\uparrow (\downarrow)}, \label{currentEq}
\end{equation}
where $-e$ is the electron charge, $n_{\uparrow (\downarrow)}$ is
the density of spin-up (spin-down) electrons, $\sigma_{\uparrow
(\downarrow)}=en_{\uparrow (\downarrow)}\mu$ is the spin-up
(spin-down) conductivity, and the mobility $\mu$ is defined via
$\vec v_{drift}=\mu \vec E$. The spin relaxation time is labeled
with $\tau_{sf}$, and the diffusion constant with $D$. It is
assumed that the total electron density in the semiconductor is
constant, i.e. $n_{\uparrow}(x)+n_{\downarrow}(x)=N_0$.
Correspondingly, the electric field is homogeneous and coupled to
the total current density as $j=e\mu N_0E_0$.

Substituting Eq.
(\ref{currentEq}) into Eq. (\ref{contEq}) we obtain two coupled
equations for the density of spin-up and spin-down electrons:

\begin{equation}
\frac{\partial n_{\uparrow (\downarrow)}}{\partial t}=D
\frac{\partial^2 n_{\uparrow (\downarrow)}}{\partial x^2}+\mu
E_0\frac{\partial n_{\uparrow (\downarrow)}}{\partial
x}+\frac{n_{\downarrow (\uparrow)}-n_{\uparrow (\downarrow)}
}{2\tau_{sf}} \label{DEq}
\end{equation}

A steady state solution of Eq. (\ref{DEq}) can be written in the
form

\begin{eqnarray}
n_{\uparrow}=\frac{N_0}{2}-Ae^{-\alpha x}, \label{nup}  \\
n_{\downarrow}=\frac{N_0}{2}+Ae^{-\alpha x}, \label{ndown}
\end{eqnarray}
where $A$ and $\alpha$ are constants to be determined.
Substituting Eqs. (\ref{nup},\ref{ndown}) into Eq. (\ref{DEq}) we
obtain a quadratic equation for $\alpha$. The positive solution of
this equation is

\begin{equation}
\alpha=\frac{\mu E_0+\sqrt{\mu^2 E_0^2+4\frac{D}{\tau_{sf}}}}{2D},
\end{equation}
which is the inverse of the up-stream spin diffusion length
defined in Ref. \onlinecite{Flatte}.

The constant $A$ can be found from the boundary conditions imposed
on the current. We consider the case when current is unpolarized
at $x\rightarrow \infty$ and fully polarized at $x=0$:

\begin{eqnarray}
j_\uparrow(x\rightarrow \infty)=j_\downarrow(x\rightarrow
\infty)=j/2, \label{bc1} \\ j_\uparrow(x=0)=j \label{bc2} \\
j_\downarrow(x=0)=0. \label{bc3}
\end{eqnarray}
It can be easily seen that the solution (\ref{nup},\ref{ndown})
with a positive $\alpha$ automatically satisfies Eq. (\ref{bc1}).
From Eqs. (\ref{bc2},\ref{bc3}) we find

\begin{equation}
A=\frac{N_0}{\sqrt{1+4\frac{D}{\tau_{sf}\mu^2E_0^2}}-1}
\label{constA}.
\end{equation}

\begin{figure}[b]
\includegraphics[angle=270,width=8cm]{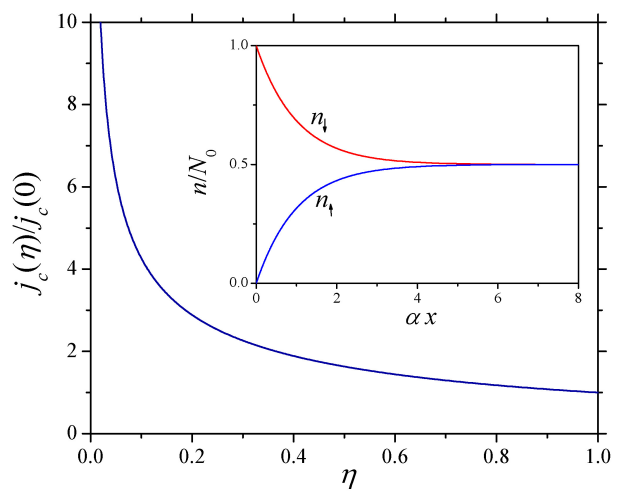}
\caption{\label{fig2} (Color online) Critical current density for
spin blockade as a function of current spin polarization $\eta$.
Inset: spin-up and spin-down densities at $j=j_c$ as a function of
the distance from the junction.}
\end{figure}

We notice from Eq. (\ref{constA}) that $A$ is a monotonically
increasing function of the current. Physically, the constant $A$ is
the deviation of the spin-up (and spin-down) electron density from
its equilibrium level at the junction. Since the maximum possible
spin polarization can only be $100\%$, the maximum possible value of
$A$ is $N_0/2$. It follows from Eq. (\ref{constA}) that the critical
current density corresponding to $A=N_0/2$ is

\begin{equation}
j_c=eN_0\sqrt{\frac{D}{2\tau_{sf}}}. \label{jc}
\end{equation}
Let us estimate this critical current density. For a GaAs structure
with $D=200$cm$^2$/s, $N_0=10^{15}$cm$^{-3}$ and $\tau_{sf}=10$ns,
the spin-down cloud extends up to about $14\mu$m at $E_0=0$, and the
critical current density for spin blockade calculated using Eq.
(\ref{jc}) is $j_c=1.7 \cdot 10^{-7}$A/$\mu$m$^2$. Such current
densities are definitely achievable in microstructures. Furthermore,
in our calculations we have used the classical noninteracting
diffusion constant $D$. Typically, due to the spin Coulomb drag
effect, the interacting diffusion constant $D_s$ is smaller
\cite{r3}.

The above analysis can be readily extended to junctions of
nonmagnetic semiconductors with ordinary ferromagnets. Let us
characterize the level of spin polarization of the current at the
junction by a parameter $\eta$ defined as

\begin{equation}
\eta=\frac{j_\uparrow(x=0)-j_\downarrow(x=0)}{j}.
\end{equation}
The limit of fully polarized spin current corresponds to $\eta=1$;
fully unpolarized to $\eta=0$. To first approximation, it can be
assumed that $\eta$ does not depend on $j$, so that the ratio
$j_\uparrow(x=0) / j_\downarrow(x=0)$ is a constant. Repeating the
above calculations we find in this case

\begin{equation}
A=\frac{\eta N_0}{\sqrt{1+4\frac{D}{\tau_{sf}\mu^2E_0^2}}-1}
\label{constA1}
\end{equation}
and
\begin{equation}
j_c=eN_0\sqrt{\frac{D}{\left(\eta^2+\eta \right) \tau_{sf}}}.
\end{equation}
Figure \ref{fig2} shows that the critical current density increases slowly by decreasing $\eta$ from $1$ to $\sim 0.3$.
Therefore, the spin blockade phenomenon is also important in
junctions with ordinary ferromagnets.

Let us now consider spin blockade in the case in which the ferromagnet has cylindrical geometry.
This analysis is relevant to STM configurations with ferromagnetic tips. A
sketch of the experimental set-up we have in mind is presented in
Fig. \ref{fig3}. Here, spin transport is studied through a
ferromagnetic tip of radius $r_1$ forming a junction with a
two-dimensional (2D) electron system. It is assumed that
spin-unpolarized electrons are injected at $r\rightarrow \infty$
and spin-up electrons are extracted at $r=r_1$. From a 2D equation
for spin density imbalance in the polar coordinates we obtain the
following expressions for spin-up and spin-down densities:

\begin{figure}[t]
\includegraphics[angle=270,width=8cm]{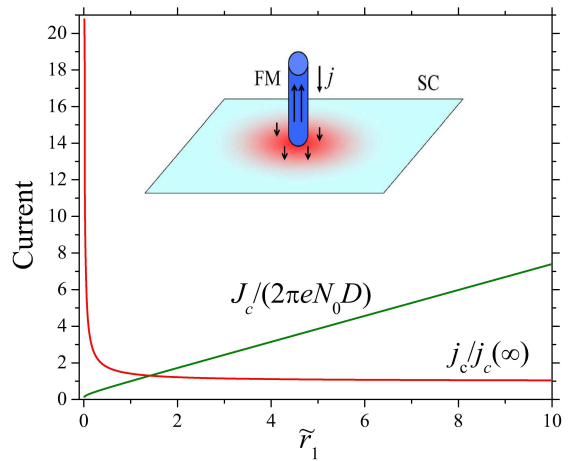}
\caption{\label{fig3} (Color online) Critical current $J_c$ and
the critical current density $j_c$ as a function of the tip radius
$r_1$. Inset: ferromagnetic tip contacting a system of electrons
constrained to 2D.}
\end{figure}

\begin{equation}
n_{\uparrow (\downarrow)}=\frac{N_0}{2}\mp
Cr^{-\frac{\gamma}{2}}K_{\frac{\gamma}{2}}\left[ \tilde r \right],
\end{equation}
where the minus sign corresponds to spin-up electrons. Here, $C$
is a constant, $\gamma=J/(2\pi e N_0 D)$, $J$ is the total
current, $K_m(x)$ is the modified Bessel function of the second
kind, and $ \tilde r=r/\sqrt{D\tau_{sf}}$. From the boundary
condition $j_\downarrow (r=r_1)=0$ we find

\begin{equation}
C=\frac{N_0 r_1^{\frac{\gamma}{2}}}{\frac{\tilde r_1
}{\gamma}\left( K_{\frac{\gamma}{2}-1}\left[
 \tilde r_1 \right]+K_{\frac{\gamma}{2}+1}\left[
 \tilde r_1 \right]
\right)-K_\frac{\gamma}{2}\left[  \tilde r_1 \right] }.
\end{equation}
Unfortunately, in the cylindrical geometry we can not derive a
closed analytical expression for the critical current from the
equation $n_\uparrow(r_1)=0$. Fig. \ref{fig3} shows a numerical
solution of this equation. The total critical current $J_c$ is
almost a linear function of $r_1$. Such dependence implies a
constant critical current density at $\tilde r=\tilde r_1$ for
$\tilde r_1\gg 1$.

We can also see that for large values of $r_1$ the critical current
density approaches the critical current density of the planar junction. Indeed, using
the asymptotic expansion $K_\nu(z) \sim e^{-z}\sqrt{\pi/(2z)}$ for
fixed $\nu$ and large $z$, the following asymptotic expression for
the critical current density at large $r_1$ is obtained

\begin{equation}
j_c \sim \frac{2}{3}eN_0\sqrt{\frac{D}{\tau_{sf}}}. \label{jcCyl}
\end{equation}
Taking into account that $2/3\approx 0.67$ and
$1/\sqrt{2}\approx0.71$, Eqs. (\ref{jc}) and (\ref{jcCyl}) are in
very good agreement.

We conclude by discussing the meaning and implications of the
spin-blockade critical current in actual experiments. The critical
current is the steady-state current that flows through the system
when the density of majority spins near the junction becomes equal
to zero. Therefore, a further increase of the current through the
junction with a fixed level of spin polarization is not possible at
all. This implies that in junctions with perfect ferromagnets
further current increase is not allowed. On the other hand, in
junctions with non-ideal ferromagnets a current increase may still
occur via a decrease of spin polarization $\eta$. Therefore, we
expect a saturation behavior of current-voltage characteristics in
junctions with perfect ferromagnets and a peculiarity (change of the
expected behavior) of current-voltage characteristics in junctions
with ordinary ferromagnets. Optical means provide an alternative way
to test this phenomenon.

%One may object that in all our considerations the junction with
%the ferromagnet was modeled through the boundary conditions on
%current components. This approximation is well suited for tunnel
%junctions, but requires more attention for direct junctions. It is
%indeed known that the presence of a ferromagnetic interface can
%produce spontaneous electron spin polarization near the interface
%\cite{r7,sham}. However, this effect is not very important for the
%predicted spin blockade, since it does not affect directly neither
%the spin-flipping time nor the diffusion coefficient entering Eq.
%(\ref{jc}). The prediction of spin polarization reversal due to
%spin blockade should thus hold for direct junctions.

Finally, there are several important spin relaxation mechanisms in
semiconductors \cite{DPandOthers,perNano}. One of them is due to
the interaction with nuclear spins \cite{perNano}. Due to electron
and nuclear spin-flip interactions, nonequilibrium electron spin
polarization results in nuclear polarization \cite{dnsp}. A
nonequilibrium nuclear spin polarization has been already observed
in S/F junctions \cite{r8, stephens}. The spin blockade regime is
interesting in this respect because of the high level of local
electron spin polarization, which should result in a strong local
nuclear spin polarization. Using a moving ferromagnetic STM tip
one may thus write a desirable nuclear spin polarization profile
in a semiconductor \cite{NSPI}. In addition, due to the very large
current densities one can achieve in nanostructures, the predicted
spin blockade may have unexpected consequences in molecular
spintronics.~\cite{MS}

In conclusion, we have predicted that the extraction of
spin-polarized electrons at S/F junctions may produce a pronounced
spin-blockade at a critical current. Only a single junction is
required to observe the spin blockade. This is an important
phenomenon since it implies that the observation of a current
saturation serves as a signature of spin polarization in a
semiconductor. This may be of value for such materials as silicon.
In a broader perspective, this phenomenon may have far-reaching
consequences in the spin control in mesoscopic and nanoscopic
devices.

This work is partly supported by the NSF Grant No. DMR-0133075. We
gratefully acknowledge useful discussions with L. Cywi{\'n}ski and
E. Rashba.

\end{document}